\documentclass[twocolumn,showpacs,preprintnumbers,amsmath,amssymb,floatfix]{revtex4}
\usepackage{epsfig}
\usepackage{dcolumn}
\usepackage{bm}
\renewcommand{\>}{\rangle}
\newcommand{\<}{\langle}
\newcommand{\rr}{{\mathbf{r}}}
\newcommand{\rp}{{\mathbf{r}'}}

\begin{document}
\title{Fusion process studied with preequilibrium giant dipole resonance  in time-dependent Hartree-Fock theory}
\author{C. Simenel$^{1,2}$, Ph. Chomaz$^{2}$ and G. de France$^{2}$}
\affiliation{$^{1}$ DSM/DAPNIA/SPhN, CEA/SACLAY, F-91191
Gif-sur-Yvette Cedex, France}  
\affiliation{$^{2}$ Grand Acc\'{e}l\'{e}rateur National d'Ions Lourds (GANIL), 
CEA/DSM-CNRS/IN2P3, Bvd Henri Becquerel, BP 55027,F-14076 CAEN Cedex 5, France} 
\date{\today}
\begin{abstract}
The equilibration of macroscopic degrees of freedom during the fusion of heavy nuclei, 
like the charge and the shape,  are studied in the Time-Dependent Hartree-Fock theory.
The preequilibrium Giant Dipole Resonance (GDR) is used to probe the fusion path.
It is shown that such isovector collective state is excited in N/Z asymmetric fusion and to a less extent 
in mass asymmetric systems. The characteristics of this GDR are 
governed by the structure of the fused system in its preequilibrium phase, 
like its deformation, rotation and vibration. 
In particular, we show that a lowering of the preequilibrium GDR energy is expected as compared 
to the statistical one. 
Revisiting experimental data, we extract an evidence of this lowering for the first time.
We also quantify 
the fusion-evaporation enhancement due to $\gamma$-ray emission from the preequilibrium GDR. 
This cooling mechanism along the fusion path may be suitable to synthesize in the future 
super heavy elements using radioactive beams with 
strong N/Z asymmetries in the entrance channel.
\end{abstract}
\maketitle

\section{Introduction}

The fusion of two nuclei occurs at small impact parameters when the overlap between their wave functions
is big enough to allow the strong interaction to overcome the Coulomb repulsion. Heavy-ion fusion reactions have
numerous applications, like the study of high spin states in yrast and super-deformed bands \cite{jan91} or the formation
of Heavy and Super Heavy Elements (SHE) \cite{hof98}. 
Induced by beams of unstable nuclei, this mechanism will also allow to produce 
very exotic species and allow for the  study of isospin equilibration in the fused system. 

The fusion process can be schematically divided in three steps: {\it (i)} an approach phase during which 
each nucleus
feels only the Coulomb field of its partner 
and which ends up when the nuclear interaction starts to dominate, {\it (ii)} a rapid equilibration of the energy 
and the angular momentum transfered from the relative motion to the internal degrees of freedom, leading to 
the formation of a Compound Nucleus (CN) and {\it (iii)} a statistical decay of the CN.
Lots of theoretical and experimental efforts \cite{das98} are made to understand step {\it (i)}. These studies  
focus on an energy range located around the fusion barrier. At these energies
the fusion is controlled by quantum tunneling 
which is strongly influenced by the couplings between the internal degrees of freedom and the relative 
motion of the two colliding partners.  
Although the cooling mechanisms involved in {\it (iii)} are well known and consist mainly in light particle
and $\gamma$-ray emission in competition with fission for heavy systems, the initial conditions of the statistical decay depend 
on the equilibration process {\it (ii)} which is still subject to many debates nowadays. Indeed, step {\it (ii)} is characterized by
an equilibration of several degrees of freedom like the shape \cite{bon80} 
or the charge \cite{bon81} which can be 
accompanied by the emission of preequilibrium particles.
Such emission decreases the excitation energy
and the angular momentum.
The latter quantities are crucial and must be determined precisely because they have a major influence 
on the CN survival probability and therefore on the synthesis of very exotic systems 
such as the SHE. 

In this paper we study the equilibration of the charges in fused systems, its interplay with other 
macroscopic degrees of freedom like the shape and the rotation, and its implications on the statistical decay.
To probe theoretically and experimentally 
this way to fusion, we use the preequilibrium isovector Giant Dipole Resonance (GDR)
\cite{cho1,bar1,bar2,bar3,sim1}. 
Giant Resonances  
are interpreted as the first quantum of collective vibrations involving 
protons and neutrons fluids. 
The Giant Monopole Resonance can be described as a breathing mode, an 
alternation of compression and dilatation of the whole nucleus. 
The GDR corresponds to a collective oscillation of the protons against the neutrons. The Giant
Quadrupole Resonance consists in a nuclear shape oscillation between prolate and oblate
deformations. Many other resonances have been
discovered \cite{Wo87,Har01}. In particular  
Giant Resonances have
been observed in hot nuclei formed by fusion \cite{New81,gaa}.
This demonstrates the survival of ordered vibrations in very excited systems,
which are known to be chaotic, even if some Giant Resonance
characteristics like the width are affected by the
temperature \cite{bra89,cho95}. Moreover, the strong couplings between various collective modes
which occur for Giant Resonances built on the ground state \cite{sim2,fal} are still present in fusion
reactions \cite{sim1,cho03}. It might therefore be possible to use the Giant Resonances properties to probe
the nuclear structure of the composite system on its way to fusion.
 
The choice of the preequilibrium GDR, that is,  a GDR excited in step {\it (ii)} before 
the formation of a fully equilibrated CN, is motivated by the fact that its 
properties strongly depend on the structure of the state on which it is built, for instance the deformation \cite{bon81}.
The idea is to form a CN with two N/Z asymmetric reactants. Such a reaction may lead 
to the excitation of a dipole mode because of the presence of a net dipole moment in the 
entrance channel. This dipole oscillation should occur {\it before} the charges are fully equilibrated, 
that is,  during the preequilibrium phase in which the system keeps a memory of the entrance channel 
\cite{bon81,cho1,sur89,bar1,bar2,bar3,das,sim1}. In addition, for such N/Z asymmetric reactions, an enhancement of the 
fast GDR $\gamma$-ray emission is expected as compared to the "slower" statistical $\gamma$-ray 
yield \cite{bar1,bar2,bar3,sim1}. This is of particular interest since the properties of these GDR $\gamma$-rays characterize  
the dinuclear system which precedes the hot equilibrated CN. The first experimental 
indications on the existence of such new
phenomenon have been reported in \cite{fli,cin,pie03fus,pie05,amo04} for fusion reactions and in 
\cite{cam95,san,pap,amo,pie03deep,pap03,amo04,pap05} in the case of deep inelastic collisions. 

The paper is organized as follows:
In Sec. II we study 
the properties of the preequilibrium GDR using the Time-Dependent Hartree-Fock (TDHF) formalism. 
In Sec. III we show how an N/Z asymmetric entrance channel may increase the fusion-evaporation 
cross-sections. Finally, we conclude in section IV.

\section{TDHF study of the preequilibrium Giant Dipole Resonance}

At the early time of the fusion reaction, the system keeps the memory of the entrance channel. 
We call this stage of the collision the {\it preequilibrium phase} which ends 
when all the degrees of freedom are equilibrated in the compound system and when the
statistical decay starts.

One of these degrees of freedom is the isospin, which measures the asymmetry between
protons and neutrons.
When the two nuclei have different N/Z ratios, the proton 
and neutron centers of mass of the total system do not coincide.
As shown in \cite{cho1,das}, there is a non zero force 
between the two kind of nucleons which tends to restore the initial isospin asymmetry.
In such a case, an oscillation of protons against neutrons on the way to fusion might occur, 
that is,  the so-called preequilibrium GDR \cite{bon81,sur89,cho1,bar1,bar2,bar3,sim1}. 

In fusion reactions the shape of the system changes drastically during the preequilibrium phase. 
Studies of the dynamics in the fusion reaction mechanism requires 
sophisticated calculations to extract the preequilibrium GDR characteristics 
(energy, width...) and in turn, on the way to fusion. To achieve this goal, we choose 
to use, as in the pioneer work of Bonche and Ng\^o on charge equilibration \cite{bon81}, 
the TDHF approach because it is a fully microscopic theory 
which takes into account the quantal nature of the single particle dynamics.
Moreover in the present study we will restrict ourself to the observation of one-body observables
(e.g. the density $\rho (r)$) which are supposed to be well described 
by such a mean field approach.
However it is clear that an important challenge is to develop methods going beyond mean field
which is beyond the scope of this paper.

In this section we present quantum calculations on preequilibrium giant collective 
vibrations using the TDHF theory. We shall start with a brief 
description of the TDHF theory in Sec. \ref{subsec:tdhf}. Then we examine the role of 
various relevant symmetries in the entrance channel, namely the N/Z and mass symmetries
 (Sec. \ref{delta}-\ref{couplings}). 
Finally, in Sec. \ref{compar} we shall compare our results with the experimental data obtained by 
Flibotte {\it et al.} \cite{fli}.

\subsection{TDHF approach \label{subsec:tdhf}}

In the TDHF approach \cite{har,foc,dir30,vau,eng75,bon,neg}, 
each single particle wave function is propagated 
 in the mean field generated by the
ensemble of particles. The mean field approximation does not take into account
the dissipation due to
two-body interactions \cite{gon,won,lac,jui}.
However TDHF takes care of one-body mechanisms
such as Landau spreading and evaporation damping \cite{cho2}. 
Quantum effects 
induced by the single particle dynamics like shell effects or modification of the moment of inertia \cite{sim04}
are accounted for properly.

The main advantage of TDHF is its fully microscopic treatment of the N-body dynamics
with the same effective interaction as the one used for the calculation of the Hartree-Fock (HF) ground sates of the collision partners.
The consistency of the method for the structure of nuclei and the nuclear reactions 
increases its prediction power and its availability to study the interplay between exotic structures and reaction mechanisms.

Moreover the TDHF equation is strongly non linear
which is of great importance for reactions around the barrier because it includes couplings between 
relative motion and internal degrees of freedom of the collision partners.
Also TDHF provides a good description of collective motion and can even 
exhibit couplings between collective modes \cite{sim2}.
In fact the TDHF theory is optimized for the prediction of expectation values
of one-body observables
and gives their exact evolution in the extreme case 
where the residual interaction vanishes.
However, the TDHF prediction of multipole moments in nuclear collision, 
for instance, may differ from the correct evolution 
because of the omission of the residual interaction.
An improvement of the description would be given by
the inclusion of the effect of the residual interaction on the dynamics, 
which would increase considerably the computational time 
and is beyond the scope of this paper.

The TDHF theory describes the evolution of the one-body density matrix $\rho (t)$
of matrix elements $\<\rr sq|{\hat{\rho}}|\rp s'q'\>=\sum_i \, \varphi_i^*(\rp s'q') \, \varphi_i(\rr sq)$,
where  $ \varphi_i(\rr sq) = \<{\mathbf{r}}sq|i\>$ denotes the
component with a spin $s$ and isospin $q$ of the occupied single particle
wave-function $\varphi_i$. 
This evolution is determined by a non linear Liouville-von Neumann equation, 
\begin{equation}
i\hbar \frac{\partial }{\partial t}{\rho} -\left[{h}(\rho ),{\rho} \right]=0
\label{eq:tdhf}
\end{equation}
where ${h}(\rho )$ is the matrix associated to the self consistent mean-field Hamiltonian. 
We have used the code built 
by P. Bonche and coworkers \cite{kim} with an effective Skyrme interaction  \cite{sky56} 
and SLy4$d$ parameters  \cite{kim}. In its actual version, TDHF does not account for pairing interactions.

\subsection{N/Z asymmetric reactions \label{delta}}

As far as the dipole motion in the preequilibrium phase is concerned, 
it is obvious that the main relevant asymmetry responsible for such a motion 
is a difference in the charge-to-mass ratio between the collision partners \cite{cho1}.
The associated experimental signature is an enhancement of the $\gamma$-ray emission
in the GDR energy region of the compound system \cite{fli,cin,pie03fus,pie05,amo04} 
which is attributed to a dipole oscillation.
Several informations about the fusion path can be extracted from such a dipole oscillation 
and its corresponding $\gamma$-ray spectrum. 
For numerical tractability we start our study of the fusion process 
with a light 
system:
$^{12}$Be+$^{28}$S$\rightarrow ^{40}$Ca.
We first deduce the $\gamma$-ray spectrum from the dipole motion. 
Then we study the effects of the deformation of the compound system, 
and of the impact parameter on this motion.

\subsubsection{The preequilibrium GDR $\gamma$-ray spectrum \label{GDRspec}}

We first consider a central  collision at an energy of $1$~MeV/nucleon in the center of mass.
The  expectation value of the dipole moment $\hat{Q}_D$ is defined 
by 
\begin{equation}
Q_D = \<\hat{Q}_D\> = \frac{NZ}{A} (X_p-X_n)
\label{eq:qd}
\end{equation}
where 
$X_p = \sum_p \frac{\<\hat x_p\>}{Z}$ 
and
$X_n = \sum_n \frac{\<\hat x_n\>}{N}$ are the positions of the proton and neutron centers 
of mass respectively. 
 The expectation value of the conjugated dipole moment $\hat{P}_D$ is then associated to the relative 
velocity between protons and neutrons, and is defined by the relation
\begin{equation}
P_D = \<\hat{P}_D\> = \frac{A}{2NZ}(P_p-P_n)
\label{eq:pd}
\end{equation}
where 
$P_p = \sum_p \<\hat p_p\>$
and 
$P_n = \sum_n \<\hat p_n\>$
are the total proton and neutron moments respectively.
These definitions ensure the canonical  commutation relation $\left[\hat{Q}_D,\hat{P}_D\right]=i\hbar$.

\begin{figure}
\begin{center}
\epsfig{figure=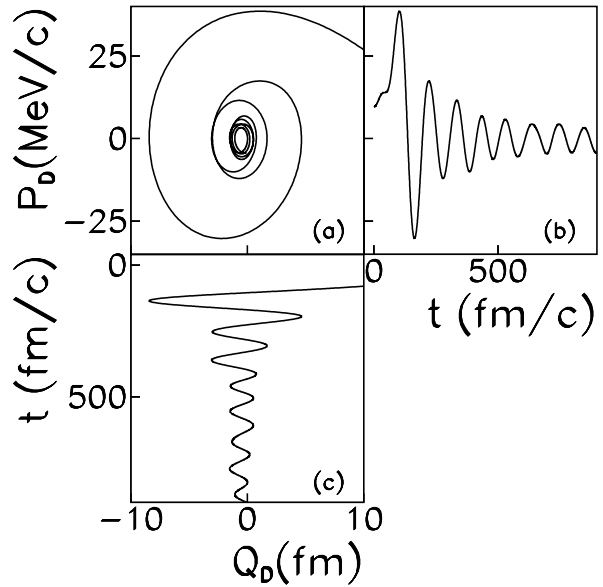,width=8cm}
\caption{
Time evolution of the expectation value of the dipole moment, $Q_D$,  and its conjugated moment, $P_D$, 
in the reaction $^{12}$Be+$^{28}$S$\rightarrow ^{40}$Ca 
at an energy of 1 MeV/nucleon in the center of mass  and at zero impact parameter.}
\label{asym}
\end{center}
\end{figure}

The time evolutions of $Q_D$ and $P_D$ are plotted in Figs.~\ref{asym}-c and \ref{asym}-b respectively. 
The trajectories in both the ($Q_D$,t) and ($P_D$,t) planes exhibit oscillations which we  
attribute to the preequilibrium GDR.
We also note that $P_D(t)$ oscillates in phase quadrature with $Q_D(t)$ and that those 
oscillations are damped due to 
the one-body dissipation.
Consequently, the plot of $P_D$ as a function of $Q_D$ shown in Fig.~\ref{asym}-a is a spiral.
The GDR period extracted from these plots is around $107$ fm/c, which corresponds to an 
energy of $\sim 11.6$ MeV.

\begin{figure}
\begin{center}
\epsfig{figure=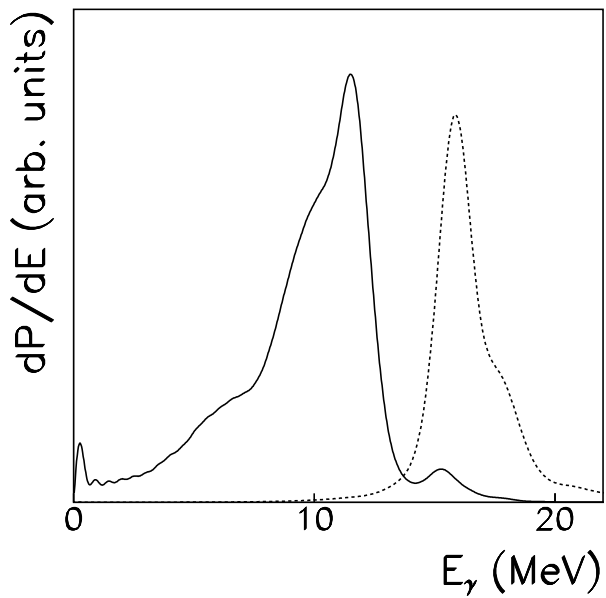,width=8cm}
\caption{preequilibrium GDR $\gamma$-ray spectrum calculated in the reaction 
$^{12}$Be+$^{28}$S $\rightarrow$ $^{40}$Ca (solid line) at an energy of 1 MeV/nucleon
in the center of mass
and $\gamma$-ray spectrum of a GDR built on the ground state of 
 $^{40}$Ca (dotted line).}
\label{spectre}
\end{center}
\end{figure}

During the collision and before the equilibrium is reached, a fast rearrangement 
of charges occurs within the composite system \cite{bon81}, generating the $\gamma$-ray emission. 
We extract the preequilibrium GDR $\gamma$-ray spectrum from the Fourier 
transform of the acceleration of the charges \cite{jac,bar3}
\begin{equation}
\frac{dP}{dE_\gamma}(E_\gamma)=\frac{2\alpha}{3\pi}\frac{|I(E_\gamma)|^2}{E_\gamma}
\label{fourier}
\end{equation}
where $\alpha$ is the fine structure constant and
$$I(E_\gamma)=\frac{1}{c}\int_0^\infty \!\!\! dt \,\, \frac{d^2Q_D}{dt^2}\exp\left(i\frac{E_\gamma t}{\hbar}\right).$$
The spectrum obtained from Eq.~\ref{fourier} is plotted in Fig.~\ref{spectre} (solid line).
In order to have a spectrum without spurious peaks coming from the finite 
integration time, we multiply the quantity $\frac{d^2Q_D}{dt^2}$ by a gaussian 
function $\exp\left(-\frac{1}{2}(\frac{t}{\tau})^2\right)$ \cite{mar05}. 
In addition, this function plays a role of a filter in the time domain.
This filter prevents the signal to be affected by the interaction
between the nucleus and the emitted nucleons which have
been reflected on the box  \cite{rei06}. 
We choose $\tau = 320$ fm/c in our calculations.
This ensures the fact that the spectra are free of spurious 
effects coming from the echo. However this procedure
adds a width $\Gamma\sim \frac{\hbar}{\tau} \sim 0.6$ MeV.
This is a drawback if one is concerned with detailed spectroscopy. 
However, in this paper, we are only interested by
the gross properties of the preequilibrium GDR
in order to study the fusion mechanisms.
%
As we can see in Fig.~\ref{spectre}, the 
preequilibrium GDR energy is $E^p_{GDR} = 11.64$ MeV, which corresponds to the previous value 
deduced from the GDR oscillation period. 

The energy of the preequilibrium GDR is much lower than the one of the GDR built on the spherical ground state of the $^{40}$Ca. This situation will be now explored into more details.

\subsubsection{Deformation effect \label{defor}}

\begin{figure}
\begin{center}
\epsfig{figure=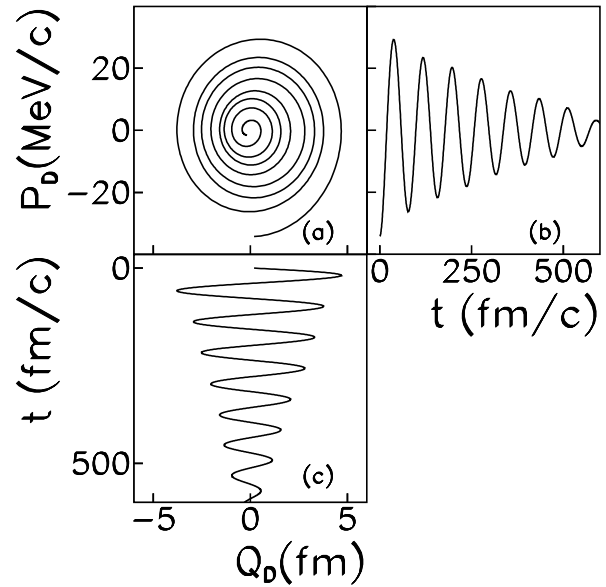,width=8cm}
\caption{
GDR built upon the HF ground state in $^{40}$Ca and excited by an isovector dipole boost:
evolution of the expectation value of the associated dipole moment, $Q_D$, and its conjugated moment, $P_D$, 
as a function of time.}
\label{ca}
\end{center}
\end{figure}

\begin{figure}
\begin{center}
\epsfig{figure=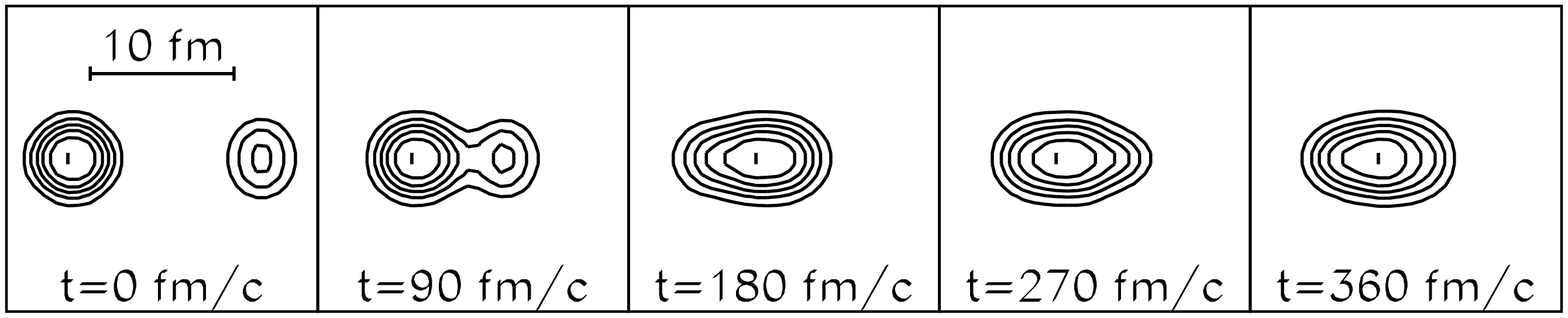,width=8.5cm}
\caption{Density plots projected on the reaction plane for different times in the case of the fusion reaction. 
Lines represent isodensities.}
\label{dens}
\end{center}
\end{figure}

To better characterize the preequilibrium GDR, it is necessary to compare it with the
usual GDR built upon the CN ground state \cite{sur89}. 
This GDR is generated by applying 
an isovector dipole boost with a velocity $k_D$ on the $^{40}$Ca HF ground 
state $|\psi(t)\> = \exp\left({-ik_D\hat Q_D}\right)|HF\>$ yielding 
an oscillation of $Q_D(t)$ and $P_D(t)$ in phase quadrature as we can see 
in Fig.~\ref{ca}. The period of the oscillation is around 80 fm/c which is lower 
than in the fusion case and corresponds to a higher energy ($E_{GDR}=15.5$ MeV) as it is shown 
in the associated  GDR $\gamma$-ray spectrum in Fig.~\ref{spectre} (dotted line). 
The lower energy obtained for the fusing system reveals a strong prolate deformation 
\cite{sim1,bon81,sur89,bar3}. The two mechanisms (fusion reaction and dipole boost) 
are expected to generate 
a GDR with quite different dynamical properties. This can be seen in the density plot projected in 
the reaction plane shown in Fig.~\ref{dens}, which shows that in the case of a fusion reaction, 
the CN relaxes its initial prolate elongation along the collision axis with a 
time which is larger than the typical dipole oscillation period of the 
GDR.

Deformation effects can be studied all along the fusion path \cite{bon80,sur89}. 
The quadrupole 
deformation parameter $\epsilon$ is defined by a scaling 
of the axis from a spherical to 
a deformed shape along the $x$-axis
\begin{eqnarray}
R_x & = & R_0(1+\alpha) \nonumber \\
R_{yz} & = & R_0(1-\epsilon)  
\label{def_epsilon}
\end{eqnarray}
where $\alpha$ is defined by the conservation of the volume of the nucleus $R_xR_yR_z=R_0^3$, which leads to 
\begin{equation}
\alpha = \frac{(2-\epsilon)\epsilon}{(1-\epsilon)^2}\, \, \, .
\label{eq:alpha}
\end{equation}
If one neglects high order terms in $\epsilon$, 
we get the usual value $\alpha \simeq 2\epsilon$.

The deformation parameter is related to the expectation values of the monopole and quadrupole 
moments $\hat{Q}_{0}$ and $\hat{Q}_{2}$ which are expressed by
\begin{eqnarray}
Q_{0} =\<\hat{Q}_{0}\> & = & \frac{1}{\sqrt{4\pi}} \int \!\!\! d{\mathbf r} \,\,  \rho({\mathbf r}) r^2 \label{mono} \\
Q_{2} =\<\hat{Q}_{2}\> & = & \sqrt{\frac{5}{16\pi}}\int \!\!\! d{\mathbf r} \,\, \rho({\mathbf r}) r^2  \left(3\frac{x^2}{r^2}-1\right).
\label{momexpres}
\end{eqnarray}
We can write $Q_{2}$ as a function of $Q_{0}$
\begin{equation}
Q_{2}= -\frac{\sqrt{5}}{2}Q_{0}+3\sqrt{\frac{5}{16\pi}} \int \!\!\! d{\mathbf r} \,\, \rho({\mathbf r}) x^2.
\label{eq:Q2Q0}
\end{equation}
Eqs.~\ref{def_epsilon} and \ref{mono} lead to
\begin{equation}
\int \!\!\! d{\mathbf r} \,\, \rho({\mathbf r}) x^2 = (1+\alpha)^2 \frac{\sqrt{4\pi}}{3} Q_{0}.
\label{eq:rhoQ0}
\end{equation}
Using Eqs.~\ref{eq:alpha}, \ref{eq:Q2Q0}, \ref{eq:rhoQ0} and $\epsilon<1$, we get
\begin{equation}
\epsilon(t)=1-\left(1+\frac{2Q_2(t)}{\sqrt{5}Q_0(t)}\right)^{-\frac{1}{4}}
\label{eq:def}
\end{equation}
which, at first order in $\epsilon$, becomes
\begin{equation}
\epsilon(t) = \frac{Q_{2}(t)}{2\sqrt{5}Q_{0}(t)}.
\label{eq:def2}
\end{equation}

\begin{figure}
\begin{center}
\epsfig{figure=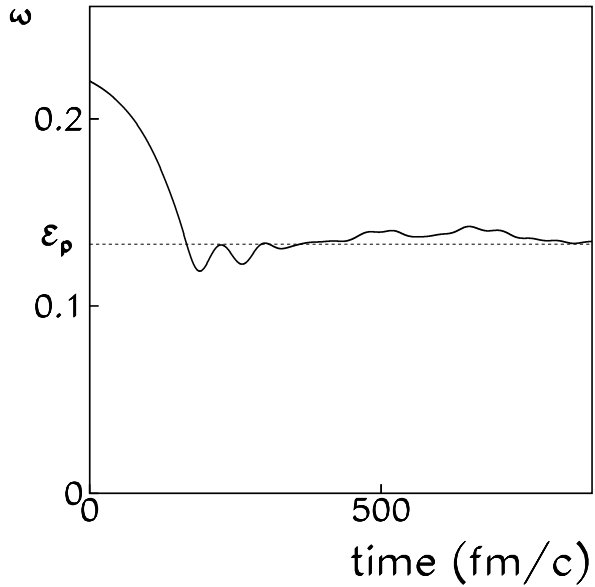,width=8cm}
\caption{Time evolution of the deformation, $\epsilon$, in $^{40}$Ca 
formed in  the $^{12}$Be+$^{28}$S fusion reaction 
at an energy of 1 MeV/nucleon in the center of mass.
The time axis origin is chosen when the maximum of the fusion barrier is reached.
The average preequilibrium deformation $\epsilon_p$ obtained from
the GDR energy (see Eq.~\ref{eq:defGDR}) is represented by a dashed line.
}
\label{deformation}
\end{center}
\end{figure}

In Ref. \cite{sim1} we used Eq.~\ref{eq:def2} to characterize the average deformation.
In Fig.~\ref{deformation} we present the time evolution of the deformation, $\epsilon(t)$,
obtained from the more general expression of $\varepsilon$ given in Eq.~\ref{eq:def}. 
We consider a $^{40}$Ca formed in the $^{12}$Be+$^{28}$S fusion reaction at an energy of 1 MeV/nucleon
in the center of mass. 
The important point here is that the deformation does not relax and strongly affects the frequency of 
the oscillations. A lower energy is expected for the longitudinal collective 
motion $ E^p_{GDR}$ in the fused system as compared to the one simulated in a 
spherical $^{40}$Ca \cite{bon81,sur89,bar1,bar2,bar3,sim1}. Following a macroscopic model for the dipole oscillation, 
we expect the energy of the GDR to evolve with the deformation along 
the $x$-axis (collision axis) as 
\begin{equation}
\frac{E^p_{GDR}}{E_{GDR}}  =  \frac{R_0}{R_x}= (1-\epsilon_p)^2
\label{eq:defGDR}
\end{equation}
where $\epsilon_p$ is the average deformation during the preequilibrium stage. 
The frequency of the GDR along the deformation axis fulfills this relation 
with $\epsilon_p\simeq 0.13$ in excellent agreement with the observed deformation 
in Fig.~\ref{deformation}.

\begin{figure}
\begin{center}
\epsfig{figure=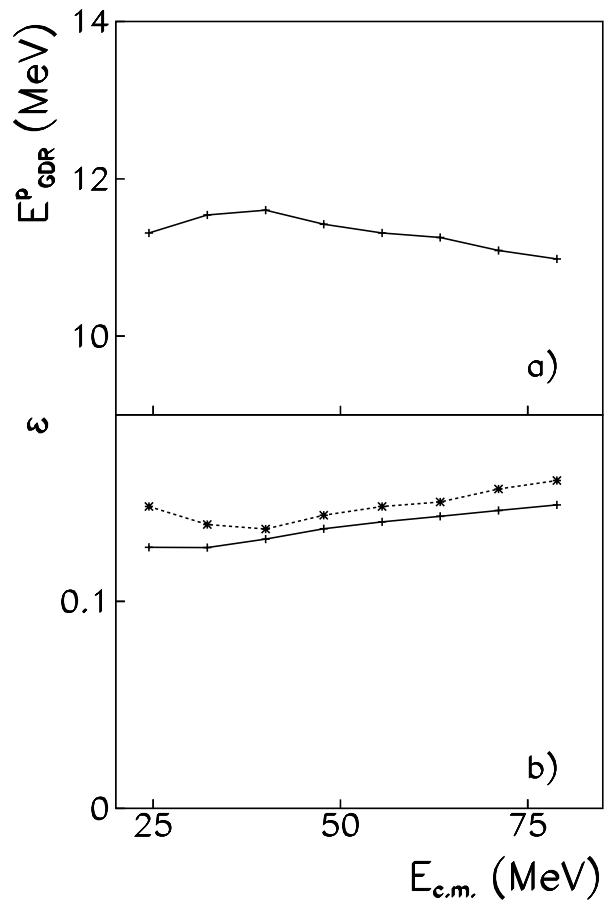,width=8cm}
\caption{{\it a)} Energy of the preequilibrium GDR obtained from the first oscillation of the dipole moment 
and {\it b)} the deformation parameter, $\epsilon$, obtained from Eq.~\ref{eq:defGDR} (dashed line) and from 
Eq.~\ref{eq:def} (solid line), as a function of the center of mass energy.}
\label{energy}
\end{center}
\end{figure}

We have also investigated the effect of the center of mass energy $E_{CM}$ on the 
preequilibrium GDR energy and on the deformation parameter (see Fig.~\ref{energy}). 
The GDR energy exhibits small 
variations (less than 1 MeV) with the center of mass energy (Fig.~\ref{energy}-a). 
For $E_{CM}<40$ MeV, the increase of $E^p_{GDR}$ with $E_{CM}$ is attributed to the formation 
of a dinuclear system with a slow neck dynamics at low energy \cite{bar3}. 
The presence of the neck is in fact expected to slow down the charge equilibration 
process, and then to increase the GDR period. 

For $E_{CM}>40$ MeV, 
Fig.~\ref{energy}-a a decrease of $E^p_{GDR}$ 
when $E_{CM}$ increases. As illustrated in Fig.~\ref{energy}-b, this is associated to a larger quadrupole 
deformation when the collision is more violent. 
Consequently, the higher the center of mass energy, 
the more prolately deformed the CN. 
In Fig.~\ref{energy}-b, the deformation is estimated from 
Eq.~\ref{eq:defGDR} (dashed line) and from 
Eq.~\ref{eq:def} (solid line) at the first maximum 
after one oscillation of $\epsilon(t)$ 
(e.g. at $t\sim 225$ fm/c in the case of $E_{CM}=1$ MeV/u 
as we can see in Fig.~\ref{deformation}). 
We also observe in this energy domain a good agreement 
between the deformations calculated with both methods.

This lowering of the GDR energy due to 
deformation is not specific to nuclear physics. Indeed, an energy splitting 
of the isovector dipole mode has been observed in fissioning atomic clusters 
due to a strong prolate deformation of the fission phase \cite{cal}.
In such systems, the use of LASERs with the "pulse and probe" technique
is expected to give access to the deformation and also to the fission time \cite{din05}.

\subsubsection{non central collisions}

\begin{figure}
\begin{center}
\epsfig{figure=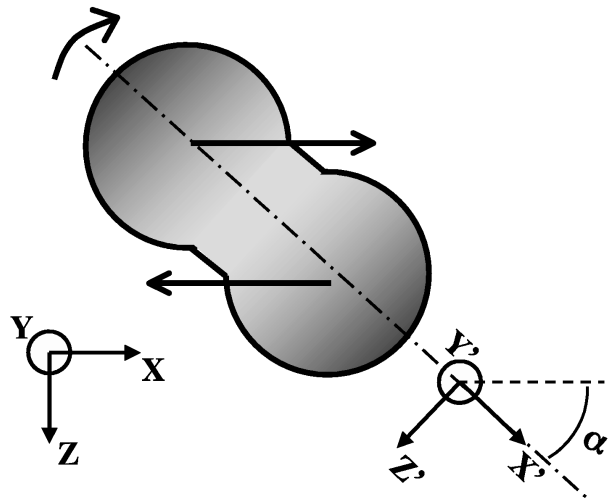,width=8cm}
\caption{Description of the two frames used in non central collisions.}
\label{coordinates}
\end{center}
\end{figure}

To better mimic the situation of a fusion reaction, we extended our calculations to
non-zero impact parameters. In fact, a non central collision may 
excite collective rotational states in the deformed preequilibrated CN. 
This rotation may be coupled to the preequilibrium GDR \cite{sur89}. In particular, the interplay of 
dipole vibration and deformation can be affected by the rotation. In addition to the 
center of mass coordinates with $x$ along the beam axis and $y$ perpendicular to the 
reaction plane, we define a new coordinate system $x'$, $y'$, $z'$, where $x'$ is the 
deformation axis, and $y=y'$ is the rotation axis (see Fig.~\ref{coordinates}). In 
the head-on collision example studied previously, those two frames are the same. 
For symmetry reasons, the dipole oscillation cannot occur along the $z=z'$ and $y=y'$ axis. 

\begin{figure}
\begin{center}
\epsfig{figure=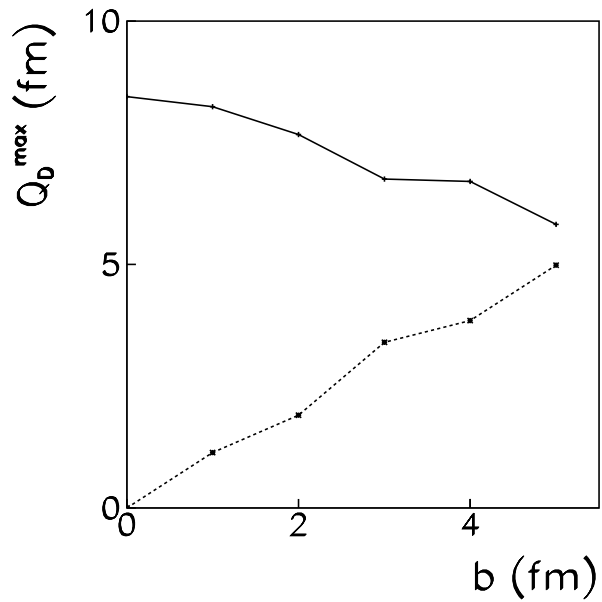,width=8cm}
\caption{Amplitude of the first oscillation of the dipole moment along $x'$ (solid line) and along $z'$ (dashed line) 
as a function of the impact parameter, $b$, in the $^{12}$Be+$^{28}$S fusion reaction 
at an energy of 1 MeV/nucleon in the center of mass.}
\label{param}
\end{center}
\end{figure}

For non-central collisions, the oscillation is only forbidden along the  $y=y'$ axis \cite{bon81,sur89}. 
In this case the amplitude of the oscillation along $x'$ slightly decreases with the 
impact parameter. This decrease becomes significant at rather large impact parameters as we 
can see in Fig.~\ref{param} where we have plotted the amplitude of the first oscillation 
of the dipole moment along $x'$ (solid line) as a function of the impact parameter. 
This decrease is accompanied by an oscillation of the dipole moment along the $z'$ axis  
with a smaller amplitude which increases with the impact parameter $b$. 
Both amplitudes are of the same order when $b\sim 5$ fm.

The oscillation along the $z'$ axis results from a weak symmetry breaking due to the 
rotation of the system \cite{sim1}. In order to demonstrate this, let us start with the time-dependent 
Schr\"odinger equation in the laboratory frame $\mathcal{R}$:
$ i\hbar  |\dot{\psi}\> = \hat{H} |\psi\>.$
In the rotating frame $\mathcal{R}'$, the expression of the wave function  is 
$ |\psi '\> = \hat{R}(\alpha) |\psi\>$ where $\hat{R}(\alpha) = e^{-i\alpha(t)\hat{J}_y}$ 
is a rotation matrix, $\hat{J}_y$ is the generator of the rotations around $y$ and $\alpha(t)$ 
is the angle between the two frames (see Fig.~\ref{coordinates}). We express the 
Schr\"odinger equation as
$- \hbar \dot{\alpha} \hat{J}_y \hat{R}^{-1}|\psi '\> + i \hbar \hat{R}^{-1} |\dot{\psi} '\> = \hat{H} \hat{R}^{-1}|\psi '\> $ 
and we get \cite{sim1}
\begin{equation}
i\hbar |\dot{\psi}'\> =\left(\hat{R}\hat{H}\hat{R}^{-1}+\hbar \dot{\alpha}\hat{J}_y\right)|\psi '\>.
\label{schrod}
\end{equation}

Eq.~\ref{schrod} is the Schr\"odinger equation expressed 
in the rotating frame $\mathcal{R}'$ of the CN  
and $\hat{H}' = \hat{R}\hat{H}\hat{R}^{-1}+\hbar \dot{\alpha}\hat{J}_y$ 
is the Hamiltonian expressed in this frame.
The last term induces a motion along the $z'$ axis from a dipole vibration along $x'$. 
It is quantified by the dipole moment along $z'$ which is plotted 
as a dashed line in Fig.~\ref{param}.  
This is a clear manifestation of couplings between rotational and vibrational motions in nuclei.

In this subsection we have shown that an N/Z asymmetry in the entrance 
channel generates a dipole oscillation during the preequilibrium phase of a fusion reaction. 
In the next one we will see that, due to polarization effects, such a motion also occurs 
in N/Z symmetric systems although with a smaller amplitude.

\subsection{N/Z symmetric reactions \label{noDelta}}

\begin{figure}
\begin{center}
\epsfig{figure=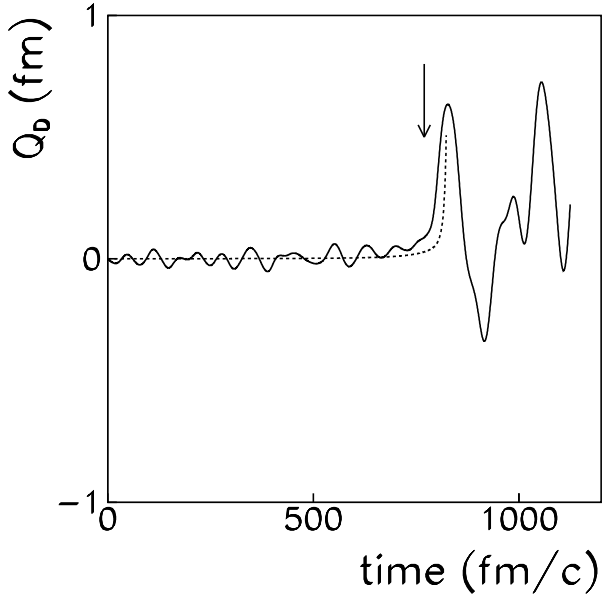,width=8cm}
\caption{Time evolution of the total dipole moment for the $^{8}$Be+$^{32}$S$\rightarrow^{40}$Ca reaction 
at an energy of 1 MeV/nucleon in the center of mass. At time $t=0$ fm/c, the distance between the centers of mass of the nuclei is 92.8 fm. The arrow indicates the time when the fusion barrier is reached. The dashed line gives the result of the adiabatic model (cf. Eq.~\ref{eq:dnp}).}
\label{sym}
\end{center}
\end{figure} 

We now examine the situation of a central collision involving 
two $N = Z$ nuclei using the example of $^8$Be+$^{32}$S at $E_{CM}=1$ MeV/nucleon
($^8$Be is bound with a strong prolate deformation in Hartree-Fock calculations with 
the SLy4$d$ force). As we can see in Fig.~\ref{sym}, 
the amplitude of the dipole oscillations is significantly reduced as compared to the 
N/Z asymmetric case (cf. Fig.~\ref{asym}-c). In this latter system ($^{12}$Be+$^{28}$S), 
the dipole oscillations are generated by the N/Z asymmetry, whereas in the 
$^{8}$Be+$^{32}$S reaction, they are only due to the mass 
asymmetry of the two collision partners. 
Indeed, a mass asymmetry induces a 
difference in the isovector polarization in the collision partners. 
This polarization is
 due to Coulomb repulsion between protons of 
the colliding nuclei {\it before} the fusion starts \cite{bon81}.

\begin{figure}
\begin{center}
\epsfig{figure=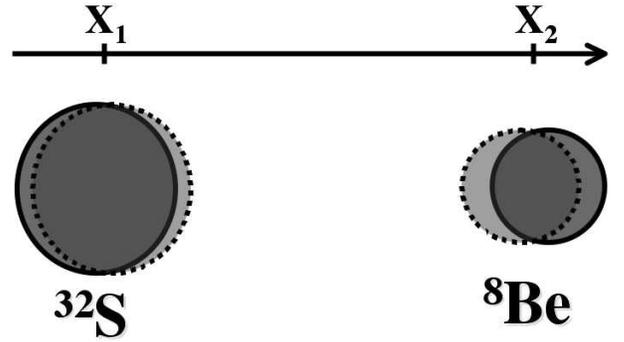,width=8cm}
\caption{Schematic representation of the isovector polarization due to Coulomb repulsion 
between protons that occurs before fusion. The protons are represented by a solid line and the neutrons by a dotted line.  $X_i$ is the position of the center of mass of the nucleus $i$.}
\label{fig:polar}
\end{center}
\end{figure} 

To show it, let us use an adiabatic approach in which we consider that the 
polarization of a nucleus at a distance $X=X_2-X_1$ between the centers of mass 
is generated by the Coulomb field of its collision partner. 
$X_i$ is the position of the center of mass of the nucleus $i$.
The distance between the proton and neutron centers 
of mass in nucleus $i$ is supposed to be small as 
compared to $X$ (see Fig.~\ref{fig:polar}). 
The equality between the external Coulomb field and 
the restoring force between protons and neutrons  leads 
to a dipole moment in the nucleus $i$
$$Q_{D_i}(t) \simeq (-1)^i \frac{N_i Z_i Z_j e^2\hbar^2}{A_i {E_{GDR}}_i^2 m X(t)^2}$$
where $i\neq j = 1$ (for $^{32}$S) or 2 (for $^{8}$Be).
The GDR energy is calculated in each collision partner 
from the dipole response frequency following a small amplitude dipole boost.
We get $E_{GDR}=23.0$ MeV for $^{32}$S and $E_{GDR_x}=17.2$  MeV
for $^{8}$Be along its deformation axis which is chosen to be aligned 
with the collision axis.
The dipole moment in the total system becomes
\begin{equation}
Q_D(t) = \frac{N_1Z_2-N_2Z_1}{A}X(t)+Q_{D_1}(t)+Q_{D_2}(t).
\label{eq:dnp}
\end{equation}
The first term of the right hand side of Eq. \ref{eq:dnp} is usually dominant
for a N/Z asymmetric reaction \cite{bar2}.
However, it vanishes for a N/Z symmetric one.
In this case, one is left with the sum of the dipole moments of the partners.
This simple adiabatic model (dashed line in Fig.~\ref{sym}) 
gives the good trend of the total dipole moment up to the vicinity of the contact point.

 After the fusion starts, the dipole moment increases and oscillates in the 
preequilibrium system. The adiabatic model is too simple 
to describe this phenomenon.
In fact, due to the polarization, 
the nuclear interaction acts first on neutrons and then is expected to modify 
strongly the dipole moment at the initial stage of the fusion \cite{bon81}. 

The consequence of this polarization in a mass asymmetric system is a
dipole oscillation which can be interpreted, as previously, in term of an excitation 
of a preequilibrium GDR. However, the GDR excitation is very small 
as compared to the N/Z asymmetric case.
Of course, for a mass and N/Z symmetric reaction no preequilibrium GDR are 
allowed for symmetry reason \cite{sim1}. 
As we will see in the next section, the special case of an N/Z asymmetric and
mass symmetric system exhibits some interesting behaviors as far as the collective
motions are concerned.

\subsection{mass asymmetry and isoscalar vibrations \label{couplings}}

In this subsection we study the couplings between the isovector dipole motion 
and isoscalar vibrations in the preequilibrium phase and their dependence on 
the mass asymmetry in the entrance channel.
The dipole motion can be coupled to isoscalar vibrations through the non linearity
of the TDHF equation \cite{sim2,sim1,cho03}.
The presence of such isoscalar vibrations in the preequilibrium system depends on the
structure of the colliding partners and on their mass asymmetry.
For instance a mass symmetric system has a stronger quadrupole deformation 
at the touching point than a mass asymmetric one. In such a system a quadrupole vibration
might appear. 

\begin{figure}
\begin{center}
\epsfig{figure=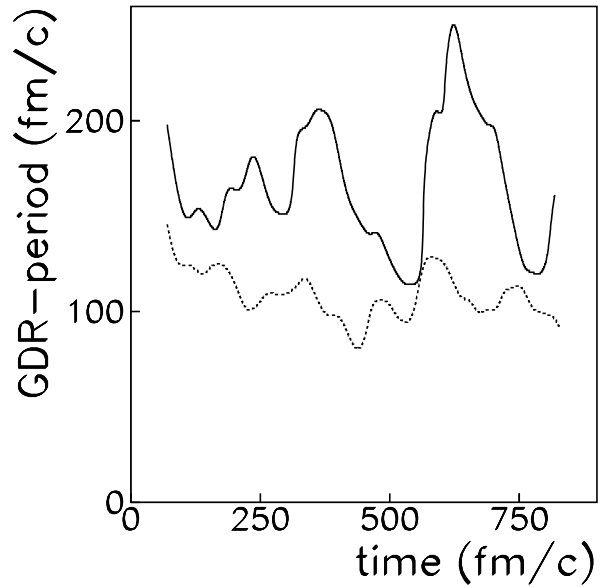,width=8cm}
\caption{Time evolution of the GDR period for $^{20}$O+$^{20}$Mg at 1.6 MeV/nucleon
(solid line) and for $^{12}$Be+$^{28}$S at 1~MeV/nucleon (dashed line).
Both energies are in the center of mass.}
\label{periode}
\end{center}
\end{figure}

Let us start this study with the time evolution of the instantaneous dipole period 
\cite{sim1} which is very sensitive to couplings with isoscalar vibrations. 
We define this period as being  
twice the time to describe half a revolution in the spiral diagram representing the evolution
of the system in the $(P_D,Q_D)$ space. 
The resulting evolution is plotted in Fig.~\ref{periode} for two N/Z 
asymmetric central collisions:
\begin{itemize}
\item the mass asymmetric $^{12}$Be+$^{28}$S reaction at $E_{CM}=1$~MeV/nucleon.
\item the mass symmetric $^{20}$O + $^{20}$Mg  
reaction at $E_{CM}=1.6$~MeV/nucleon.
\end{itemize}
The center of mass energy has been chosen to obtain the same  
$E_{CM}/V_B$ ratio
for both reactions ($V_B$ is the Coulomb barrier).

The mean values of the GDR period obtained for the two
reactions are different. For the mass symmetric reaction, this value 
is $\simeq 170$ fm/c, whereas in the mass asymmetric case it is 
$\simeq 105$ fm/c, in good agreement with the one obtained from 
Fig.~\ref{asym} (107 fm/c). This difference is attributed to a larger 
deformation of the CN in the mass symmetric case which, in average, 
is $\epsilon \sim 0.2$ (from Eq. \ref{eq:def2}), as compared to the mass asymmetric 
system ($\epsilon \sim 0.13)$. 
Note that it is not appropriate to use Eq.~\ref{eq:defGDR}, to calculate 
the deformation from the observed GDR energy frequency for $^{20}$O+$^{20}$Mg 
since it is valid only for small deformations.

\begin{figure}
\begin{center}
\epsfig{figure=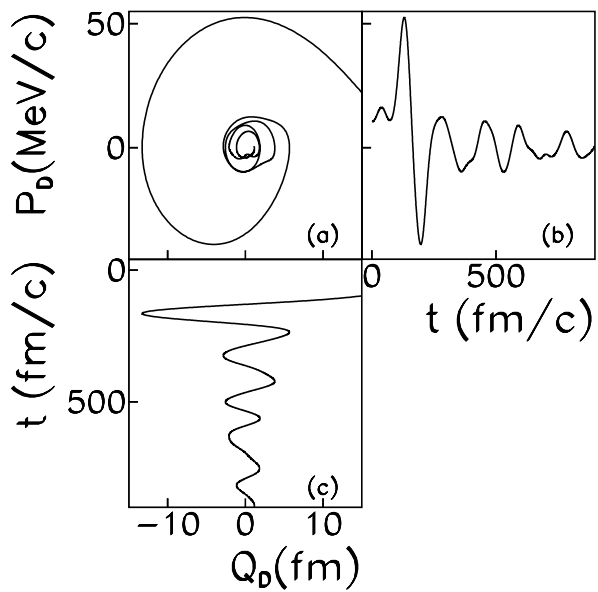,width=8cm}
\caption{Evolution of the expectation value of the dipole moment, $Q_D$, and its conjugated moment, $P_D$, 
in the reactions $^{20}$O+$^{20}$Mg$\rightarrow ^{40}$Ca at an energy of 1.6 MeV/nucleon in the 
center of mass.}
\label{spiraleOMg}
\end{center}
\end{figure}

\begin{figure}
\begin{center}
\epsfig{figure=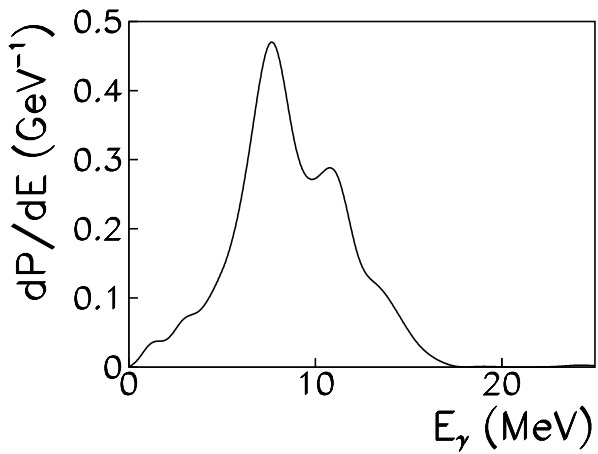,width=8cm}
\caption{GDR $\gamma$-ray spectrum calculated in the $^{20}$O+$^{20}$Mg$\rightarrow ^{40}$Ca reaction
at an energy of 1.6 MeV/nucleon in the center of mass.}
\label{OMgspectra}
\end{center}
\end{figure}

The dipole moment time evolution for those two reactions  
(Figs.~\ref{asym} and \ref{spiraleOMg}), shows that unlike $^{20}$O+$^{20}$Mg, 
the oscillations in the $^{12}$Be+$^{28}$S system 
are dominated by a single energy. This is consistent 
with the evolution of the GDR period in Fig.~\ref{periode} which is 
rather constant in the mass asymmetric case whereas it 
exhibits strong oscillations in the mass symmetric one. This anharmonicity 
can also be seen in the GDR $\gamma$-ray spectrum of the $^{20}$O+$^{20}$Mg 
reaction plotted in Fig.~\ref{OMgspectra}. Indeed, one can clearly 
identify two peaks in this spectrum at 7.7 MeV and 10.8 MeV. 

\begin{figure}
\begin{center}
\epsfig{figure=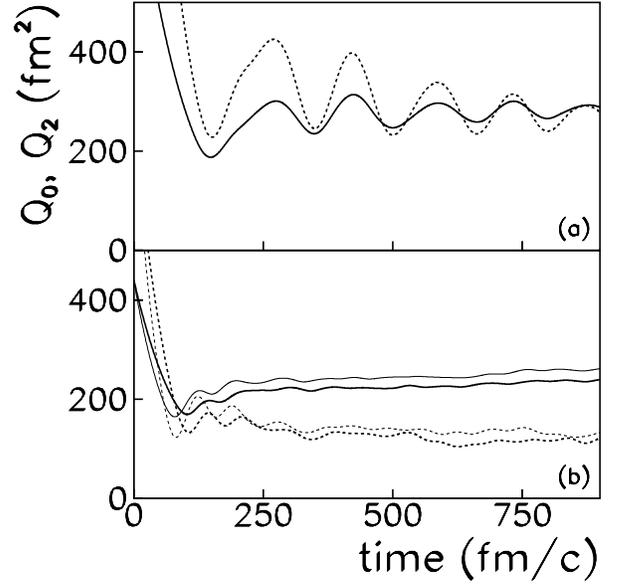,width=8cm}
\caption{Evolution with time of the monopole ($Q_0$, solid line) and quadrupole ($Q_2$, dashed line) 
moments in the reactions $^{20}$O+$^{20}$Mg$\rightarrow ^{40}$Ca at 1.6 MeV/nucleon (a) and 
for $^{12}$Be+$^{28}$S at 1 MeV/nucleon (thick lines) and 1.6 MeV/nucleon (thin lines) (b). Both energies are in the center of mass.}
\label{t_q0_q2}
\end{center}
\end{figure}

To better understand what is the origin of the differences 
between the two systems, we have 
calculated the evolutions of the monopole $Q_0$ and quadrupole $Q_2$ 
moments defined by Eqs.~\ref{mono} and \ref{momexpres} respectively. 
Those evolutions are plotted in Fig.~\ref{t_q0_q2}-a for $^{20}$O+$^{20}$Mg. 
We first note that $Q_2$ is always positive, that is, 
the compound system keeps a prolate deformation. In addition,
$Q_0$ and $Q_2$ exhibit strong oscillations with the same period  $\sim165$ fm/c. 
Therefore, we conclude that they have the same origin which 
is interpreted as a vibration of the density around a prolate shape \cite{sim1}.
This mode is only excited in the mass symmetric channel: 
the evolutions of $Q_0(t)$ and $Q_2(t)$ for the mass asymmetric reaction 
($^{12}$Be+$^{28}$S)
at 1 MeV/nucleon in the center of mass (thick lines in Fig.~\ref{t_q0_q2}-b) 
do not show any significant oscillation of these moments. 
Evolutions of $Q_0(t)$ and $Q_2(t)$ at 1.6 MeV/nucleon in the center of mass 
are also plotted (thin lines in Fig.~\ref{t_q0_q2}-b). They do not 
exhibit any significant oscillation neither. Therefore, the vibrations observed 
in Fig.~\ref{t_q0_q2}-a are not attributed to a difference in the collision energy
but to the mass asymmetry in the entrance channel.

The monopole and quadrupole oscillations
modify the properties of the dipole mode in a time dependent way
\cite{sur89,bar2}. 
Let us consider  a harmonic oscillator for the dipole motion 
with a time dependent rigidity constant. 
This is a way to simulate 
the non linearities of TDHF. Indeed, the observed oscillation of the 
density modifies the restoring force between protons and neutrons.
This is due to the fact that the density enters in the 
mean field potential of the TDHF equation (Eq.~\ref{eq:tdhf}). This restoring 
force is lower along the deformation axis of a prolately deformed 
nucleus than in the perpendicular axis. Thus, variations of the density 
profile in the TDHF equation can be modeled by a corresponding variation 
of the rigidity constant $k(t)$. In such a model, 
the evolution of the dipole moment is given 
by the differential equation $ \ddot Q_D(t)+(k(t)/\mu) Q_D(t)=0$ where 
$\mu=\frac{NZ}{A}m$ is the reduced mass of the system. 
We note $\omega_0$ the average pulsation related 
to the  rigidity constant given by $k(t)/ \mu=\omega_0^2(1+\eta \cos{\omega t})$, 
where $\omega$ is the pulsation of the density oscillation deduced from
Fig.~\ref{t_q0_q2}-a and $\eta$ is a  
dimensionless constant which quantifies the coupling between the GDR and 
the other collective mode associated to the density vibration. We thus have
\begin{equation}
\ddot Q_D(t)+\omega_0^2[1+\eta \cos{\omega t}]Q_D(t) = 0.
\label{eq:mathieu1}
\end{equation}
This equation is the so called Mathieu's equation \cite{sim1}. 

It is interesting to show how we can get this equation from a more microscopic equation like the TDHF one 
(Eq.~\ref{eq:tdhf}) in a  one dimensional framework.  
Following the way of ref. \cite{sch}, the Wigner transform of Eq.~\ref{eq:tdhf} 
for a local self consistent potential $V$ is
\begin{equation}
\frac{\partial f }{\partial t} + \frac{p}{m} \frac{\partial f}{\partial x} = \frac{2}{\hbar}\sin\left(\frac{\hbar}{2}\frac{\partial^V }{\partial x}\frac{\partial^f }{\partial p}\right)V f
\label{WF}
\end{equation}
where $f(x,p,t)=\int \!\! ds\,\, \exp(-ip.s/\hbar)\,\rho(x\!+\!\frac{s}{2},x\!-\!\frac{s}{2},t)$ 
is the Wigner transform of the density matrix $\rho(x_1,x_2,t) = \<x_1|\hat{\rho}(t)|x_2\>$. 
The upper indices on the derivative operators in Eq.~\ref{WF} stand for the function on 
which the operator acts. We have of course  $f = f_p+f_n$ where $f_p$ 
and $f_n$ are the Wigner transforms of the proton and neutron density matrices respectively.

We now apply the Wigner Function Moment (WFM) method to get a closed system 
of dynamical equations for the dipole and its conjugated moments. 
We calculate the integrals on the phase space of Eq.~\ref{WF} with 
the weights $x \tau$ on the one hand, and $p \tau$ on the other hand ($\tau\!\!=\!\!1$ 
for protons and $-1$ for neutrons). The distance $D$ between proton and neutron centers of mass can be written as
$D=\int \!\! dx \, dp\,\, x\, (f_p -f_n)$ and we get 
$$
\dot D + \int \!\!\! dp \,\,\frac{p}{m} \,\int\!\!\! dx\, \,x \,\frac{\partial}{\partial x} \,(f_p - f_n) $$
$$
= \frac{2}{\hbar} \int \!\!\! dx\,dp\,\, x \,\sin\left(\frac{\hbar}{2}\frac{\partial^V }{\partial x}\frac{\partial^f }{\partial p}\right)V \,(f_p - f_n)
$$
where the time dependence has been omitted for simplicity.
The right hand side term is the integral of multiple $p$-derivatives of $f$ so it vanishes because $f_p$, $f_n$ and all their $p-$derivatives 
vanish for $|p|\rightarrow \infty$. With $P$ being the relative momentum between 
protons and neutrons $P = \int  dp\, dx \,p \,(f_p-f_n)$ we get
\begin{equation}
\dot D = \frac{P}{m}.
\label{int1}
\end{equation}

We now calculate the integral of Eq.~\ref{WF} with the weight $p\tau$. Noting 
the matter density $n(x,t) = \int dp \,f(x,p,t)$ and the kinetic energy density 
${\mathcal A}(x,t) = \frac{1}{m} \int dp\, p^2 \,f(x,p,t)$ we have
$$
\dot{P}+\int \!\!\! dx \,\, \frac{\partial}{\partial x} \left({\mathcal A}_p-{\mathcal A}_n\right)=-\int \!\!\! dx \,\, \frac{\partial V}{\partial x}\left(n_p-n_n\right).
$$
Using ${\mathcal A}=0$ for $|x| \rightarrow \infty$ we have
\begin{equation}
\dot{P} = -\int \!\!\! dx \,\, \frac{\partial V}{\partial x}\left(n_p-n_n\right).
\label{int2}
\end{equation}

Eqs.~\ref{int1} and \ref{int2} are the system of dynamical equations of 
motion we were looking for. It is important to stress that this system of equations 
is obtained without approximation for a local potential. To go further, we need 
an explicit form of the potential. If we consider for instance a harmonic oscillator 
$V=kx^2/2$, we obtain the dipole moment evolution equation: 
$ m\ddot{D} = -k\, D$ with the solution $D = D_0 \cos{\omega_0 \,t}$, where $\omega_0=\sqrt{k/m}$.

If a breathing mode occurs at a pulsation $\omega$, then the density $n(x,t)$ oscillates with the pulsation $\omega$: 
$n(x,t) = n_0(x)\left[1+\lambda(x)\cos{\omega t}\right]$.
Since the potential is self consistent, it also presents oscillations which are a function of $\cos{\omega t}$: $V(x,t)\equiv V(x,\cos{\omega t})$.
We assume for this potential the separable form 
$V(x,t)=V_0(x)\left(1+{\mathcal F}\left[{\cos{\omega t}}\right]\right)$,
where $V_0(x)$ is the potential when no breathing mode is excited.
Using a harmonic picture for $V_0$, that is,  $V_0(x) = \frac{1}{2}m\omega_0 x^2$,
we get from Eqs.~\ref{int1} and \ref{int2} the equation for the dipole moment 
$Q_D = \frac{NZ}{A}D$:
\begin{equation}
\ddot{Q}_D(t)+\omega_0^2\left(1+{\mathcal F}[{\cos{\omega t}}]\right)Q_D(t) = 0.
\label{WTresult}
\end{equation}
We finally see that the Mathieu's equation (Eq.~\ref{eq:mathieu1}) appears 
to be an approximation of Eq.~\ref{WTresult} where only the linear part of 
the function ${\mathcal F}(\xi)\simeq \eta \xi$ is conserved.

\begin{figure}
\begin{center}
\epsfig{figure=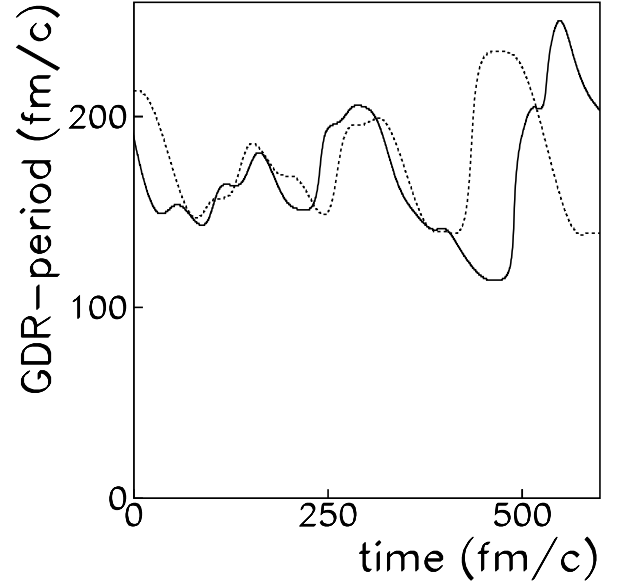,width=8cm}
\caption{Time evolution of the GDR period calculated for the reaction $^{20}$O+$^{20}$Mg$\rightarrow$ $^{40}$Ca
at an energy of 1.6 MeV/nucleon in the center of mass
(solid line) and its modelization by the Mathieu's equation (dashed line).}
\label{mathieu}
\end{center}
\end{figure}

We have solved the Mathieu's equation numerically with 
a set of parameters suitable for our problem.  
The pulsation of the density oscillation is extracted from Fig.~\ref{t_q0_q2}-a and we get $\omega \simeq 7.5$ MeV$/\hbar$. 
For the pulsation of the GDR we choose the main peak at $\omega_{GDR}\simeq 7.7$ MeV/$\hbar$ (see Fig.~\ref{spiraleOMg}).
It is related to the pulsation $\omega_0$ by the relation $\omega_0 = r \omega_{GDR}$. 
The constants $r$ and $\eta$ are tuned to reproduce approximatively 
the TDHF results period. 
The parameter $r$ is expected to be close to $1$ but not exactly $1$ 
because of the presence of the oscillating term which may slightly change the 
mean value of the dipole pulsation. The solution of the Mathieu's equation 
oscillates with a time-dependent period which reproduces the TDHF case quite well 
with $r \simeq 1.1$ and $\eta\simeq 0.5$ (see Fig.~\ref{mathieu}). 

In a recent paper \cite{cho03}, following the formalism developed 
in a study of non linear vibrations \cite{sim2}, we related $\eta$ 
to a matrix element of the residual interaction coupling collective states.

As a consequence, the excitation of collective modes such as the quadrupole
and monopole vibrations 
is coupled to the preequilibrium GDR. Such vibrations occur only in the 
mass symmetric reaction we studied. The effects of this coupling are a reduction 
of the GDR energy (estimated around 10 per cent in this case) and an additional 
spreading of the resonance line shape due to the modulation of the dipole frequency.

\subsection{comparison with experiments \label{compar}}

\begin{figure}
\begin{center}
\epsfig{figure=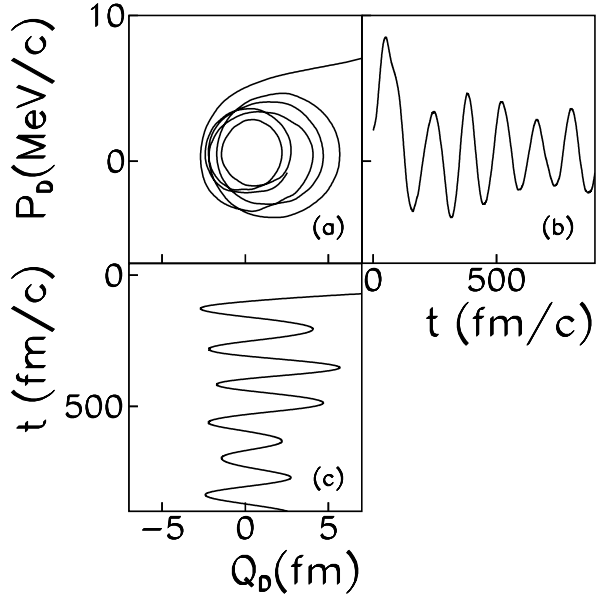,width=8cm}
\caption{Evolution of the expectation value of the dipole moment, $Q_D$, and its conjugated moment, $P_D$, 
in the case of the N/Z asymmetric reaction 
$^{40}$Ca+$^{100}$Mo at a center of mass energy of 0.83 MeV/nucleon.}
\label{spiral_fli_asym}
\end{center}
\end{figure}

\begin{figure}
\begin{center}
\epsfig{figure=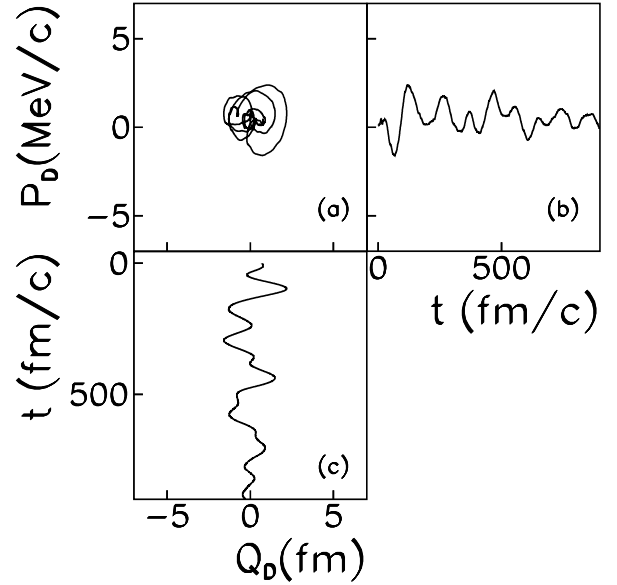,width=8cm}
\caption{Evolution of the expectation value of the dipole moment, $Q_D$, and its conjugated moment, $P_D$, 
in the case of the N/Z quasi-symmetric reaction 
  $^{36}$S+$^{104}$Pd at a center of mass energy of 0.83 MeV/nucleon.}
\label{spiral_fli_sym}
\end{center}
\end{figure}

As a test case, we have performed TDHF calculations of the reactions studied 
by Flibotte {\it et al.} \cite{fli}. In this paper, two systems have been investigated: 
an N/Z asymmetric one ($^{40}$Ca+$^{100}$Mo) and an N/Z quasi-symmetric one
($^{36}$S+$^{104}$Pd) at a center of mass energy of 0.83~MeV/nucleon. 
These systems have been chosen because they lead to the
same composite system ($^{140}$Sm). 
The corresponding dipole evolutions obtained from TDHF
are plotted in Fig.~\ref{spiral_fli_asym} for the N/Z asymmetric reaction and 
in Fig.~\ref{spiral_fli_sym} for the N/Z quasi-symmetric one. 
A dipole oscillation is observed in both reactions but with a stronger amplitude 
in the N/Z asymmetric one.

\begin{figure}
\begin{center}
\epsfig{figure=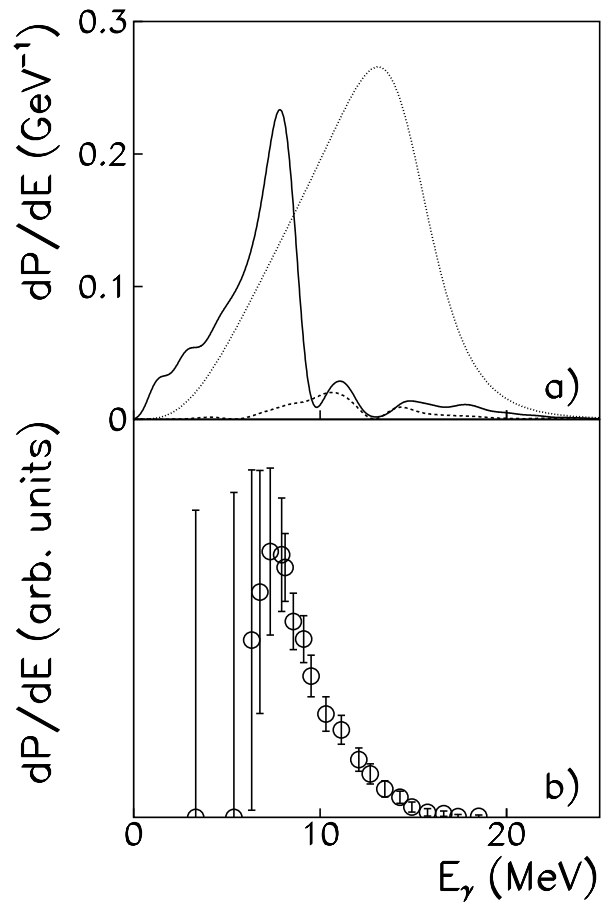,width=8cm}
\caption{
{\it a) } preequilibrium GDR $\gamma$-ray spectrum calculated in the reactions 
$^{40}$Ca+$^{100}$Mo  (solid line) and $^{36}$S+$^{104}$Pd (dashed line). 
The dotted line represents the first chance statistical $\gamma$-ray decay spectrum.
{\it b) } Experimental data resulting from the subtraction
of the $\gamma$-ray spectra obtained by Flibotte {\it et al.} \cite{fli} 
in the reactions $^{40}$Ca+$^{100}$Mo and $^{36}$S+$^{104}$Pd.
}
\label{fourier_fli}
\end{center}
\end{figure}

The preequilibrium GDR $\gamma$-ray spectra for those reactions are
calculated using Eq.~\ref{fourier} and plotted in Fig.~\ref{fourier_fli}-a. 
The area under the peak associated to the N/Z asymmetric reaction (solid line) is considerably 
larger than the one under the N/Z quasi-symmetric one (dashed line). 

To estimate the importance of the preequilibrium $\gamma$-ray emission 
with respect to the statistical decay and its role on the fusion process, we have calculated the spectrum 
associated to the first chance statistical $\gamma$-ray decay. 
It is obtained from the $\gamma$-ray emission probability in all directions per energy 
unit assuming an equilibrated CN \cite{sno,bar3,bri}. Its expression is
\begin{equation}
\frac{dP}{dE_\gamma} = \frac{4\alpha}{\pi m c^2}\frac{ \Gamma_{GDR}}{ \Gamma_{CN}}\frac{NZ}{A}\frac{E_\gamma^4\,\, e^{-\frac{E_\gamma}{T}}}{\left(E_\gamma^2-E_{GDR}^2\right)^2+\Gamma_{GDR}^2E_\gamma^2}
\label{stat}
\end{equation}
where $m$ is the nucleon mass, $\Gamma_{GDR}$ and $E_{GDR}$ are the width and the energy of 
the statistical GDR respectively, and $T$ is the temperature of the equilibrated CN. At first order, the 
energy of the GDR does not depend on the temperature \cite{gaa}.
We use the values $E_{GDR}=15$ MeV and $\Gamma_{GDR}=7$ MeV. 
Following the same method as the one  employed in Ref. \cite{bar3}, we
 approximate the CN width $\Gamma_{CN}$ with the total neutron width
\begin{equation}
\Gamma_{CN} \simeq \Gamma_n = \frac{2 m r_0^2 A^{\frac{2}{3}}}{\pi \hbar^2}T^2 e^{-\frac{B_n}{T}}
\label{eq_Gamma_n}
\end{equation}
where $B_n = 8.5$ MeV is the neutron binding energy and $r_0=1.2$ fm.
The temperature $T$ is calculated from the equation
\begin{equation}
T=\sqrt{\frac{E^*}{aA}}
\label{temperature}
\end{equation}
where $a \simeq 1/10$  MeV$^{-1}$ is the level density parameter 
and $E^*=71$ MeV is the excitation energy. 
The resulting spectrum is plotted in 
Fig.~\ref{fourier_fli}-a (dotted line). 

We note that the N/Z 
asymmetric preequilibrium spectrum is comparable in intensity to the first 
step statistical one.
This fact has already been pointed out by Baran {\it et al.} \cite{bar3} 
who got a similar spectrum for the N/Z asymmetric 
reaction with a semiclassical approach.

Another important conclusion which can be drawn from Fig.~\ref{fourier_fli}-a 
is the lowering of the GDR $\gamma$-ray energy for the non 
statistical part as compared to the statistical one 
which is attributed to the deformation of the nucleus (see sec.  \ref{defor}).
This phenomenon is also reported by Baran {\it et al.} \cite{bar3}. 
In fact we get from Fig.~\ref{fourier_fli}-a a position of the peak 
of about 7.5 MeV for the preequilibrium GDR while Baran {\it et al.} obtained 
$\sim$9 MeV. 
On the experimental side, 
the $\gamma$-ray spectra are dominated by a statistical background
decreasing exponentially. In addition to this background, 
the GDR creates a bump located around the GDR energy (Fig.~1 of Ref. \cite{fli}).
To get rid of the statistical background, the authors of \cite{fli} linearized
the $\gamma$-ray spectra by dividing them by a theoretical statistical background.
The resulting spectra are plotted in Fig.~2 of Ref. \cite{fli}.
This procedure is used by the authors to determine the preequilibrium to statistical ratio
for the GDR component. However it cannot be used to determine the positions in energy
of the peaks because the division by an exponential background induces 
a shift in energy which is different for both contributions 
(statistical and preequilibrium)
if they are not centered around the same energy, as expected from Fig.~\ref{fourier_fli}-a.

We modified the procedure as follows. First, we assume 
that no preequilibrium $\gamma$-ray is emitted 
in the N/Z quasi-symmetric reaction $^{36}$S+$^{104}$Pd.
We then subtract the total $\gamma$-ray spectrum 
associated to the quasi-symmetric reaction from the N/Z asymmetric one.
These two spectra are plotted in Fig.~1 of Ref. \cite{fli}.
The result of this subtraction is the preequilibrium component of the GDR
in the  reaction $^{40}$Ca+$^{100}$Mo, and is plotted in Fig.~\ref{fourier_fli}-b.
The error bars are both statistical and systematic due to the graphical extraction of the data. 
Below 5 MeV the systematic error is to high to get relevant data.
Focusing on the energy position of the preequilibrium component,
we note a good agreement between TDHF predictions and experimental data.

To conclude, we extracted from existing data, for the first time,  
an experimental observation of the lowering of the preequilibrium 
GDR predicted by our TDHF calculations.
This analysis shows that the preequilibrium GDR is, indeed, a powerful
experimental tool to study the fusion path. 
Another application of N/Z asymmetric fusion reactions 
is proposed in the next section.

\section{Fusion/Evaporation Cross Sections of heavy nuclei}

As mentioned in \cite{bar3}, the emission of a preequilibrium GDR $\gamma$-ray  
decreases the excitation energy hence the initial temperature of  
the nucleus reaching the statistical phase. The emission of preequilibrium 
particles, which can be controlled in our example by the N/Z asymmetry, is 
thus a new interesting cooling mechanism for the formation of Heavy and 
Super Heavy Elements. For such nuclei, the statistical fission 
considerably dominates the neutron emission and the survival 
probability of the CN becomes very small. 

SHE must be populated at low excitation energy. Firstly, because the 
smaller the excitation energy, 
the smaller the fission probability. Secondly, because the shell corrections decrease 
with excitation energy \cite{ign79}. These corrections are responsible 
for the stability of the transfermiums nuclei ($Z>100$) in their ground state. 
The quantum stabilization 
decreases quite rapidly with excitation energy until the fission barrier vanishes.
Those two reasons are strong motivations to 
study the cooling mechanisms 
involved in the preequilibrium phase of the CN formation. 

In the following, we expose one cooling mechanism responsible for the predicted 
enhancement of the survival probability in the case of a N/Z asymmetric reaction.
As an illustration, we treat only the $\gamma$-emission part of the preequilibrium 
GDR decay. Although it may play an important role, we do not treat the 
preequilibrium neutron emission for two reasons:
\begin{itemize}
\item  Only the direct neutron decay of giant resonances can be assessed 
  in TDHF. Then, we would be able to describe
   only a small part of this neutron emission, 
   the other parts being the sequential and statistical decays. 
  Missing the sequential decay would be a strong limitation of the description.
\item We would need not only the number of emitted neutrons, but also their energy.
  Consequently, huge spatial grid would have to be used in order to perform
  a spatial Fourier transform of the single particle wave functions, which is
 out of range of three dimensional TDHF codes
  because of computational limitations. 
 \end{itemize}

Let us define $P_{E^*_{init.}}(E^*)$ the survival probability at an excitation 
energy $E^*$ of a CN which started its statistical decay at the energy $E^*_{init}$. 
We also note $P_{surv}^S$ and $P_{surv}^A$ the final survival probabilities of 
the CN formed by N/Z symmetric and asymmetric reactions respectively.

\begin{figure}
\begin{center}
\epsfig{figure=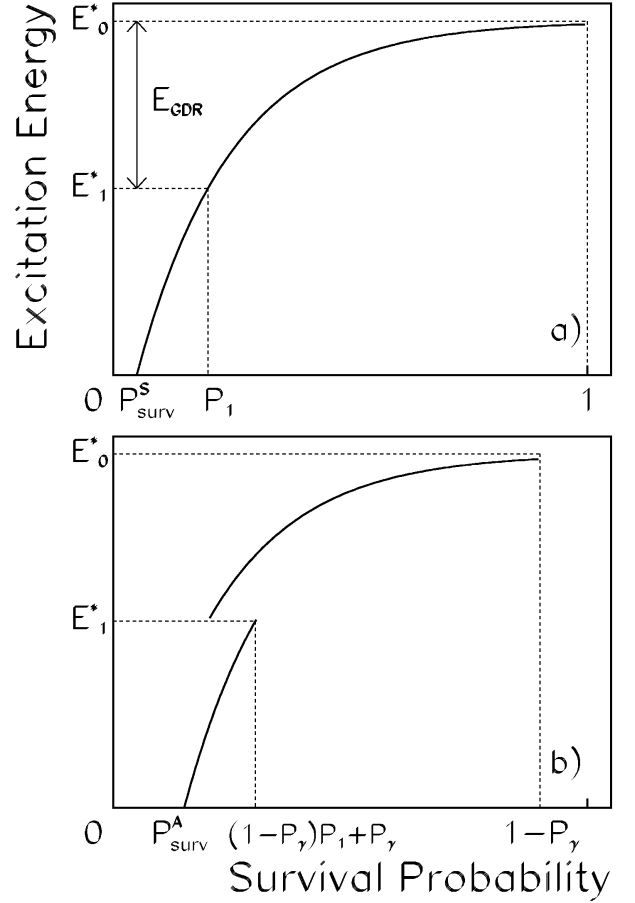,width=8cm}
\caption{Schematic representation of the CN population during the statistical decay
in the case of an N/Z symmetric collision (a) and an N/Z asymmetric reaction (b).}
\label{stat_decay}
\end{center}
\end{figure}

Fig.~\ref{stat_decay}-a  illustrates schematically the evolution of the survival 
probability ({\it x}-axis) when the excitation energy decreases ({\it y}-axis) 
in a case of an N/Z symmetry in the entrance channel. 
In this case, no $\gamma$-ray emission is expected in the preequilibrium phase 
and the initial excitation energy is always maximum $E^*_{init}=E^*_0$,
where  $E^*_0=Q+E_{cm}$ is the excitation energy when no preequilibrium particles 
are emitted, $E_{cm}$ is the center of mass energy and $Q=(M_1+M_2-M_{CN})c^2$. 
During the statistical decay, the excitation energy decreases mainly through 
neutron emission, but at the same time the survival probability of the compound nucleus
decreases too.
For instance, when the excitation energy reaches $E^*_1=E^*_0-E_{GDR}$, the survival
probability $P_1=P_{E^*_0}(E^*_1)$ at this energy might be small.
At the end of the decay, when the excitation energy is zero, the survival probability becomes 
$P_{surv}^S  =  P_{E^*_0}(0) =  P_1 P_{E^*_1}(0). $

Fig.~\ref{stat_decay}-b shows the 
same for an N/Z asymmetric reaction. In this last case, the nucleus can emit a 
preequilibrium GDR $\gamma$-ray with a probability $P_\gamma$. 
The nuclei which  
emit such a $\gamma$-ray begin the statistical decay at a lower energy $E_{init}=E^*_1$, 
whereas those which did not emit a $\gamma$-ray still starts their decay at $E_{init}=E^*_0$.
The probability for the latter case is $1-P_\gamma$.
The survival probability at the end of the decay then reads 
$P_{surv}^A =  \left[\left(1-P_\gamma\right) P_1+  P_\gamma\right] P_{E^*_1}(0)$. 
The ratio of the survival probabilities between the N/Z symmetric and asymmetric cases is
\begin{equation}
\frac{P_{surv}^A}{P_{surv}^S} =  1+ \frac{P_\gamma}{P_1} \left(1-P_1\right).
\label{enhancement}
\end{equation}

We now use a simple model to get an estimate of this quantity. 
It is clear that, to get a quantitative predictions of survival probabilities,
the studied mechanism has to be included in more elaborated statistical models,
which is beyond the scope of this paper. 
The probability $P_{\gamma}$ can be calculated by integrating Eq.~\ref{fourier} 
over the energy range. This can be 
done for example with a TDHF calculation or using the classical electrodynamic formulae 
from Ref.~\cite{jac}. Following these formulae, we approximate the probability to  
emit a preequilibrium GDR $\gamma$-ray per interval of energy by
\begin{eqnarray}
\frac{dP_\gamma}{dE} & = & \frac{2 e^2 Q_{D}(0)^2}{3\pi (\hbar c)^3}\left(E_1^2+\frac{{\Gamma_{GDR}}^2}{4}\right) \nonumber \\
 & & \frac{E_1^2 E}{\left[(E-E_1)^2+\frac{{\Gamma_{GDR}}^2}{4}\right]\left[(E+E_1)^2+\frac{{\Gamma_{GDR}}^2}{4}\right]} \nonumber
\end{eqnarray}
where  $E_1=\sqrt{E_{GDR}^2-\frac{{\Gamma_{GDR}}^2}{4}}$ 
is the ``shifted'' energy of the damped harmonic motion 
and ${\Gamma_{GDR}}$ is the damping width of the preequilibrium GDR.
The initial value of the dipole moment, $Q_D(0)$, can be estimated from Eq. \ref{eq:dnp} 
at the touching point and neglecting the polarization of the collision partners \cite{bar2}.
We get
$$ Q_D(0)\simeq \frac{R_1+R_2}{A} \left( Z_1N_2-Z_2N_1\right)$$
where $R_i$ is the radius of nucleus $i$.

To determine $P_{E^*_0}(E^*_1)$, we need to solve a system of six 
equations: Eqs.~\ref{eq_Gamma_n}, \ref{temperature} and
\begin{equation}
\frac{dE^*}{dt} = - \frac{\Gamma_n(t)}{\hbar}(B_n+T(t)) 
\label{Eex}
\end{equation}
\begin{equation}
\frac{dP}{dt} = - \frac{\Gamma_f(t)}{\hbar}P(t) 
\label{Psurv}
\end{equation}
\begin{equation}
\Gamma_f(t) = \frac{\hbar \omega_0 \omega_s}{2\pi \beta} e^{-\frac{B_f(t)}{T(t)}}
\label{eq_Gamma_f}
\end{equation}
\begin{equation}
B_f(t) \equiv B_f[E^*(t)] = B_f(0) e^{-\frac{E^*}{E_d}} 
\label{Bf}
\end{equation}
Eq.~\ref{Eex} gives the evolution of the excitation energy, assuming as in \cite{bar3}
 that the CN width
can be identified to the neutron width. 
This implies that we neglect the statistical gamma emission.
This choice is justified by the fact that the statistical neutron emission
is much more probable than the gamma emission in the excitation
energy domain of interest where the fission dominates, 
which is above the neutron emission threshold $B_n$.
Eq.~\ref{Psurv} gives the evolution of the survival probability against fission  $P$. Eq.~\ref{eq_Gamma_f} 
gives the evolution of the fission width. The parameters $\omega_0$ and $\omega_S$ are the oscillator 
frequencies of the two parabolas approximating the potential $V(x)$ in the first minimum 
and at the saddle point respectively. The variable $x$ is related to the distance between 
the mass centers of the nascent fission fragments (see \cite{ari}) and $\beta = 5\times 10^{21}$s$^{-1}$ 
is the reduced friction. Eq.~\ref{Bf} gives the evolution of the fission barrier $B_f$. 
For SHE, this barrier has only a quantum nature and vanishes at high excitation energy. 
$E_d\simeq 20$ MeV is the shell damping energy \cite{ari}. We consider that a CN with an excitation 
energy between $E^*_1$ and $E^*_0$ decays only by fission or neutron emission.

We take here the example of the reaction $^{124}$Xe+$^{141}$Xe$\rightarrow ^{265}$Hs$^*$ 
at the fusion barrier ($E_{cm}=B_{fus}$), that is, an excitation energy $E^*_0=54$ MeV. With 
an energy and a width of the GDR of 13 MeV and 4 MeV respectively, the preequilibrium $\gamma$-ray emission 
probability is $P_\gamma \simeq 0.05$. 
For the statistical decay we take $B_f[E^*=0]\simeq 8.5$ 
MeV, $B_n = 6.5$ MeV and $\omega_0 \simeq \omega_S \simeq 1$ MeV/$\hbar$. 
We also get a survival probability $P_{E^*_0}(E^*_1) \simeq 0.01$ 
which is small as compared to $P_\gamma$. 
Following Eq.~\ref{enhancement}, the enhancement of the total survival probability 
due solely to the N/Z asymmetry in the entrance channel becomes 
$P_{surv}^A/P_{surv}^S \sim 6$.

To conclude, we see that such an effect may be useful for the formation of Heavy and 
Super Heavy Elements. Indeed, based on our conclusions, very asymmetric N/Z collisions induced 
by radioactive ion beams that are coming online in several laboratories, should allow the synthesis 
SHE with a larger cross sections than are obtainable with beams of stable isotopes.

\section{conclusion}

In this paper we have performed TDHF calculations to study in some details the properties of the 
preequilibrium GDR that can be excited before the formation of a fully equilibrated CN. 
We have shown that this probe can be used to better understand 
the early stage of the fusion path, and more precisely the charge equilibration. 
We have clarified the role of the N/Z and/or mass asymmetries on the GDR excitation. 
The energy of the preequilibrium GDR is expected to decrease with excitation energy, an effect 
attributed to a strong prolate shape associated to the fused system. 
We presented the first experimental indication of this shift in energy.
The calculations for an N/Z asymmetric collisions at non zero impact parameters have been performed and  
revealed couplings between the dipole oscillations and the CN rotation. Other couplings 
between vibrational modes for mass symmetric reactions have also been studied. 

Finally we suggest that the use of N/Z asymmetric fusion reactions is a good choice to synthesize 
Heavy and Super Heavy Elements. In that case, the preequilibrium GDR $\gamma$-ray emission cooling 
mechanism might be well suited to reach the statistical phase with a low excitation energy yielding 
a larger survival probability against fission. The availability of radioactive beams with large 
N/Z asymmetry and sufficient intensities for these kind of studies will be extremely useful to 
check experimentally our predictions in the near future.

\begin{acknowledgments}

This paper is dedicated to the memory of P. Bonche, the author of the TDHF code we used. 
We are grateful to M. Di Toro for a useful reading of the manuscript.
We also  thank V. Baran, M. Colonna, D. Lacroix, D. Boilley 
and J. P. Wieleczko for several fruitful discussions, and P. Schuck for providing a pertinent reference.

\end{acknowledgments}

\end{document}